\documentclass{article}
\usepackage{graphicx}
\usepackage{fullpage}
\usepackage{amsmath,amssymb, amsfonts, amsthm, amstext,placeins, 
setspace, geometry, enumitem, hyperref, versions, tikz, tabulary, multirow,  lscape,bm,chngcntr, subcaption,caption}
\usepackage[font=small,labelfont=bf]{caption}
\usepackage[numbers]{natbib}
\usepackage{setspace}\onehalfspace
\newcommand{\E}{\mathbb{E}} 	%Expectation operator
\title{Semiparametric panel data models using neural networks}
\author{Andrew Crane-Droesch\footnote{Economic Research Service, US Department of Agriculture.  andrew.crane-droesch@ers.usda.gov.  The views expressed are those of the author and should not be attributed to the Economic Research Service or USDA}}
\begin{document}

\maketitle

\begin{abstract}
\noindent This paper presents an estimator for semiparametric models that uses a feed-forward neural network to fit the nonparametric component.  Unlike many methodologies from the machine learning literature, this approach is suitable for longitudinal/panel data.  It provides unbiased estimation of the parametric component of the model, with associated confidence intervals that have near-nominal coverage rates.    Simulations demonstrate (1) efficiency, (2) that parametric estimates are unbiased, and (3) coverage properties of estimated intervals.  An application section demonstrates the method by predicting county-level corn yield using daily weather data from the period 1981-2015, along with parametric time trends representing technological change.  The method is shown to out-perform linear methods such as OLS and ridge/lasso, as well as random forest.  The procedures described in this paper are implemented in the R package \texttt{panelNNET}.\footnote{\url{https://github.com/cranedroesch/panelNNET}}
\end{abstract}

%It is further shown that this model and estimator nests a nonparametric heterogeneous treatment effects model and estimator, which can consistently estimate individualized treatment effects conditional on covariates.

Neural networks\citep{rosenblatt_1958} (termed ``deep learning''\citep{lecun_2015} in some contexts) are the current state-of-the-art in machine learning and artificial intelligence.  They have been successfully applied to tasks ranging from computer vision, natural language processing, self-driving cars, and quantitative finance.  Hardware and software constraints limited the usefulness of these computationally-intensive approaches for many years.  However, recent advances in hardware and software -- and especially in algorithm design -- have made them increasingly accessible and effective.  

Ultimately however, neural networks are nothing more than algorithms for finding an optimal set of derived regressors from a set of input variables.  These derived regressors are then used in a linear model of some outcome.  Given that neural networks ultimately yield linear models, many standard econometric techniques can be applied to them in a fairly straightforward manner.  That insight forms the crux of this paper.

In particular, this paper presents a semi-parametric extension to cross-sectional and panel data models, using neural networks to fit the nonparametric component of the model.  The model discussed here is quite general, and is variously applicable to high-dimensional regression adjustment, heterogeneous/individualized treatment effect estimation, general nonparametric regression problems, and forecasting with longitudinal data.  

The basic model is 
\begin{equation}\label{basic}
y_{it} = \alpha_i+\bm{X}_{it}\beta + f(\bm{Z}_{it}) + \epsilon_{it}
\end{equation}

\noindent where $y$ is an outcome for unit $i$ in time $t$\footnote{The individual-time indices are presented without loss of generality.  Any context with repeated observations of one or more cross-sectional unit is admissible in this framework.}, $\bm{X}$ is a $N\times P_X$ matrix of data to be represented linearly, and $\bm{Z}$ is a $N\times P_Z$ matrix of variables to be represented nonparametrically.  Choices of which variables to be included in $\bm{X}$ or $\bm{Z}$ are left to the modeller.  As will be made clear below, it will be appropriate to include variables in $\bm{X}$ where the modeler has knowledge of appropriate parametric structure and/or desires unbiased marginal effects.  All or some elements of $\bm{Z}$ may be included in $\bm{X}$, yielding a linear ``main effect'' along with a nonlinear component.  The compound error $\alpha_i + \epsilon_{it}$ represents between-unit and within-unit variability, respectively.

Rather than estimating hundreds or thousands of individual effect parameters, standard econometric practice removes $\alpha$ via the ``within'' transformation:
\begin{equation}
y_{it}-\bar{y}_i = \left(\bm{X}_{it}-\bar{\bm{X}}_{i}\right)\beta + f^*(\bm{Z}_{it}-\bar{\bm{Z}}_{i}) + \epsilon_{it}- \bar\epsilon_{i}
\end{equation}

\noindent This fails however when $f()$ is nonlinear, because $f(\bm{Z}_{it}) \neq f^*(\bm{Z}_{it}-\bar{\bm{Z}}_{i})$.  While multidimensional linear basis expansions of $f(\bm{Z})$ can solve this problem, they are quickly overcome by the curse of dimensionality as $P_Z$ grows.  

Other machine learning approaches are also ill-suited to estimating \ref{basic}.  Random forests\citep{breiman_2001} can consistently\citep{wager_2014,wager_2015} and efficiently estimate models of the form $y = f(\bm{Z}) + \epsilon$, but they have no clear way to incorporate known parametric structure where available, nor an obvious way to eliminate the nuisance parameters $\alpha$.  Elastic-net regression\citep{zou_2005} and generalized additive models\cite{hastie_1990} can handle fixed effects by first projecting-out fixed effects, but they require extremely large multidimensional basis expansions if they are to serve as universal approximators of an arbitrary $f()$ with a high-dimensional $\bm{Z}$.  

This paper contributes to the econometric literature on nonparametric fixed effects models, which have heretofore relied on series estimation or kernel methods\citep{henderson_2008, hoderlein_2012, li_2015} and thereby suffer greatly from the curse of dimensionality.  This paper also contributes to the machine learning literature, which has not yet developed a nonparametric algorithm suited to settings with repeated observations of many cross-sectional units.  Finally, this paper contributes to the emerging literature at the intersection of machine learning and causal inference\cite{chernozhukov_2016}, particularly that which seeks to estimate personalized treatment effects\cite{wager_2015, athey_2016_trees}.

\section{Neural networks}
\subsection{Panel data}
\FloatBarrier
This paper proposes an estimator of \ref{basic} using a feed-forward neural network for the nonparametric part:
\begin{equation}\label{neuralnet}
\begin{array}{l l}
y_{it} = \alpha_i+\bm{X}_{it}\beta + \bm{V}^1_{it}\Gamma^1 + \epsilon_{it}\\
\bm{V}^1_{it} = a\left(\gamma^2 + \bm{V}^2_{it}\bm{\Gamma}^2\right)\\
\bm{V}^2_{it} = a\left(\gamma^3 + \bm{V}^3_{it}\bm{\Gamma}^3\right)\\
\hspace{1.5cm}\vdots\\
\bm{V}^L_{it} = a\left(\gamma^L + \bm{Z}_{it}\bm{\Gamma}^L\right)\\
\end{array}
\end{equation}
where $\bm{V}^l$ are derived variables or ``nodes'' at the $l$th layer.  The parameters $\bm{\Gamma}^L$ (termed ``weights'' in the computer science literature) map the data $\bm{Z}$ to the outcome via the intermediate nodes.  \textit{A priori}, no particular interpretation is attached to the derived variables $\bm{V}$ -- they are simply nonlinear combinations of $\bm{Z}$ that are chosen to make the model fit well.  

The number of layers and the number of nodes per layer (i.e.: the dimension of $\bm{V}^l$) is a hyperparameter chosen by the modeler.  Note that $\mathbf{\Gamma}^{2:L}$ are matrices of dimension equal to the number of nodes of the $l$th layer and the next layer up.  The function $a()$ is termed the ``activation'' function, and maps $V$ from the real line to a defined interval -- common choices are sigmoids such as the logistic and the hyperbolic tangent.  More recently, the ``rectified linear unit''\citep{nair_2010} (and variants \citep*{maas_2013, he_2015}) activation function -- $a(x) = max(0, x)$ -- has been shown to improve performance, especially in networks with many layers.  \citet{hornik_1989} have shown that neural networks can approximate any continuous function, given sufficiently many nodes and/or layers.

\begin{figure}[htbp]
\caption{Schematic drawing corresponding to equation \ref{neuralnet}, for a model with two parametric covariates, four nonparametric covariates, and two hidden layers with three nodes each.  Line segments represent parameters, circles represent variables or derived variables.}
\label{schematic}
\begin{center}
\includegraphics[scale = .7]{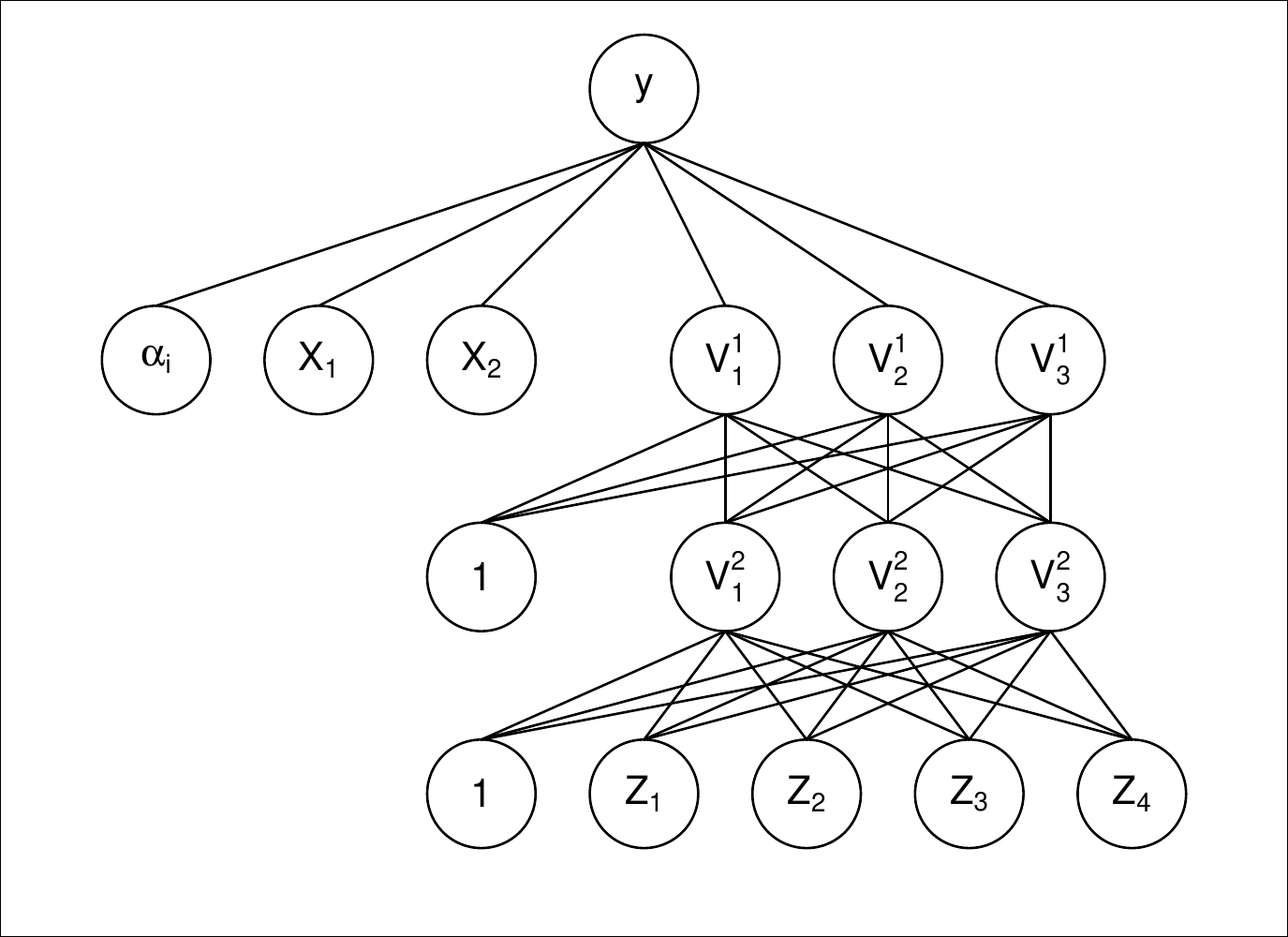}
\end{center}
\end{figure}

For a neural network with two layers, \ref{neuralnet} can be written more compactly as 
\begin{equation}\label{compact}
y_{it} = \alpha_i+\bm{X}_{it}\beta +  a(a(\bm{Z}_{it}\bm{\Gamma^3})\bm{\Gamma^2})\Gamma^1 + \epsilon_{it}
\end{equation}

Because the top layer is a linear model in the $\bm{X}$'s and the derived variables $\bm{V}^1$, the ``within'' transformation\footnote{This is without loss of generality; multi-way fixed effects can similarly be projected-out without altering the top-level parameters.} $y_{it}-\bar{y}_i = \left(\bm{X}_{it}-\bar{\bm{X}}_{i}\right)\beta + \left(\bm{V}^1_{it}-\bar{\bm{V}}^1_{i}\right)\Gamma^1 + \epsilon_{it} - \bar\epsilon_{i}$ holds without altering $\beta$ or $\Gamma^1$.  Fixed effects are thus removed.

%% \FloatBarrier
%% \subsection{Heterogeneous causal effects}
%% This model nests the heterogeneous treatment effects model.  If $D_{it}$ is some binary exogenous treatment\footnote{This is without loss of generality; $D_{it}$ may be continuous and only conditionally exogenous, and the data need not neccesarially be longitudinal.}, then the top layer of \ref{neuralnet} can be re-specified as 
%% \begin{equation}\label{HTE}
%% y_{it} = \alpha_i+\bm{X}_{it}\beta + \delta D_{it} + \bm{V}^1_{it}\Gamma^1 + \left(D_{it} \times \bm{V}^1_{it}\right)\Gamma^D + \epsilon_{it}\\
%% \end{equation}
%% 
%% \noindent leaving the lower layers unchanged.  The estimate of the causal effect thus becomes a function of covariates $\bm{Z}$ and is defined as $\hat\tau(\bm{Z}_{it}) \equiv \hat E_i\left[y_i^{D_i = 1} - y_i^{D_i = 0}| \bm{Z_i}\right] = \hat\delta + \left(\bm{V}^1_{it}\right)\hat\Gamma^D$.

\subsection{Estimation}

The above are special cases of neural networks for continuous variables, and are fit to data in the same way as the general case.  The loss surface for neural networks is generally nonconvex, and methods for minimizing this loss have been the focus of substantial research\footnote{See \citet{friedman_2001} for an introduction, and \citet{ruder_2016} for a literature review}.  Most methods are variants of minibatch gradient descent, which iteratively calculates the derivatives of the loss function with respect to the parameters for a subset of the data, alters the parameter estimates in those directions according to a step size, and proceeds iteratively with a new sample of data to convergence.  This approach is popular in large-scale predictive applications because it works with very large datasets, including those that can not fit into a computer's memory.  Quasi-Newton methods are also feasible and work well with smaller and less complex networks.  

In general, neural networks are overparameterized and will overfit the data, leading to poor performance when generalizing to new data.  One way of controlling this is to tune the number of nodes and layers.  Alternatively, the number of nodes and layers can be chosen to be sufficiently high as to ensure a good fit, and the parameters then chosen to minimize the penalized loss function $R = N^{-1}\displaystyle\sum_i (y_i - \hat{y_i}) + \lambda\displaystyle\sum \theta^2$ where $\theta$ is a vector of each of the parameters in the model, from each layer.  Because regularization induces bias, parameters for which unbiased estimates are desired should not be included in $\theta$ -- this may apply to the coefficients associated with parametric terms $\bm{X}$, if unbiased estimates of marginal effects are desired.  The tuning parameter $\lambda$ can then be chosen by some variant of cross-validation, including optimization against a withheld test set where $k$-fold cross-validation is impractical due to computational constraints.

\subsubsection{The ``OLS trick''}

While gradient descent methods generally perform well, they are inexact.  The top level of a neural network is a linear model however, in derived regressors in the typical context or in a mixture of derived regressors and parametric terms in our semiparametric context.  Ordinary least squares provides a closed-form solution for the parameter vector that minimizes the unpenalized loss function, given a set of derived regressors, while ridge regression provides an equivalent solution for the penalized case.  

In particular, after the $m$th iteration, the top level of the network is 
\[
y_{it} = \alpha_i+\bm{X}_{it}\beta^m + \bm{V}^{m}_{it}\Gamma^{m} + \epsilon_{it}\\
\]
Because gradient descent is inexact, the parameter sub-vector $[\beta^m, \Gamma^m] = \theta^m$ does not satisfy 
\begin{equation}\label{ridge}
\underset{\theta}{min}\left(\bm{y}^{dm} - \bm{W}^{dm}\theta\right)^T\left(\bm{y}^{dm} - \bm{W}^{dm}\theta\right) + \tilde\lambda\theta^T\theta
\end{equation}
\noindent where $\bm{W} = [X, V]$, $dm$ indicates the ``within'' transformation and $\tilde\lambda > \lambda$ is the penalty corresponding to the ``budget'' that is ``left over'' after fitting the lower level parameters, which generate $\bm{V}$.  Given that the general ridge regression solution is equivalent to minimizing $(\bm{(y - W\theta)^T(y - W\theta)} \text{ such that } \bm{\theta^T\theta} < c$, one may calculate the implicit $\tilde\lambda$ for the top level of the neural network by minimizing 
\[
\underset{\tilde\lambda}{min}\left(\mathcal{B}^T\mathcal{B} - \theta^{mT}\theta^{m}\right)^2
\]
where 
\[
\mathcal{B} = \bm{(W^TW}+\tilde\lambda I)^{-1}\bm{W}^T\bm{y}
\]

One may then replace $\theta^m$ with $\mathcal{B}$.  Doing so ensures that the sum of the squared parameters at the top level of the network remains unchanged, but ensures that the (top level of the) penalized loss function reaches its minimum subject to that constraint.  This facilitates approximate inference, discussed in section \ref{inference_section}.

\subsection{Inference}\label{inference_section}

Bootstrap approaches to inference will generally be computationally prohibitive, but approximate inference via linear taylor expansion is feasible\cite{rivals_2000}.  One first computes the estimate of the Jacobian $\bm\xi$, either numerically or from analytical derivatives.  For example, a neural net with two hidden layers has derivatives
\[\begin{array}{l}
\frac{\partial\hat y}{\partial \hat\beta} = \bm{X}\\
\frac{\partial\hat y}{\partial \hat\Gamma^1_{p1}} = \bm{V}^1_{p1}\\
\frac{\partial\hat y}{\partial \hat\Gamma^2_{p2}} = \displaystyle\sum_{p1} a'(\bm{V}^2\bm{\Gamma}^2)\Gamma^1_{p1}\circ\bm{V}_{p2}^2\\
\frac{\partial\hat y}{\partial \hat\Gamma^3_{p3}} = a'(\bm{V}^2\bm{\Gamma}^2)\Gamma^1\circ\displaystyle\sum_{p2} a'(\bm{Z\Gamma}^3)\Gamma^2_{p2} \circ Z_{p3}
\end{array}
\]
\noindent where $a'()$ is the derivative of the activation function and the sums are row-wise sums of concatenated column vectors.  %

Given the Jacobian $\bm{\hat\xi} \equiv \left[\frac{\partial\hat y}{\partial \hat\beta}, \frac{\partial\hat y}{\partial \hat{\Gamma}^1}, \frac{\partial\hat y}{\partial \hat{\Gamma}^2}, \frac{\partial\hat y}{\partial \hat{\Gamma}^3}\right]$, one may use it in a manner analogous to a data matrix in an OLS regression to compute the parameter covariance matrix.  Assuming homoskedasticity, this would be $\frac{\hat\epsilon^T\hat\epsilon}{N-p_\xi}\bm{\left(\hat\xi^T\hat\xi\right)}^{-1}$.  In panel-data settings, the ``cluster-robust'' estimator\citep{cameron_2012} may be preferred: 
\[
\bm{\hat{V}_\theta} = \frac{G}{G-1}\frac{N}{N-p_\xi}\bm{\left(\hat\xi^T\hat\xi+ \lambda\bm{I^*}\right)}^{-1}\displaystyle\sum_g\left(\bm{\hat\xi}^T_g\hat\epsilon_g^2\hat\epsilon_g^{2T}\bm{\hat\xi}_g\right)\bm{\left(\hat\xi^T\hat\xi+ \lambda\bm{I^*}\right)}^{-1}
\]

\noindent where $G$ is the number of clusters.  If causal inference is desired on parametric terms, $\bm{I^*} = \bm{I}$ except for those diagonal entries corresponding to the parametric terms, which should be set to zero.\footnote{Note that unbiased estimates of marginal effects of parametric terms are only guaranteed when such terms are either exogenous, or when controlling for similarly unpenalized covariates renders them conditionally independent.  If $\E[X^T\epsilon|Z] = \bm{0}$ and $Z$ enters the model in some form that is subject to penalization, then such controls will not generally solve the endogeneity problem. }  

Naturally, penalization induces bias when viewed from a frequentist perspective.  Given however that we are not interested in performing inference on penalized elements of $\bm{\hat\theta}$ itself, but rather on functions of $\bm{\hat\theta}$ (i.e.: predictions from the fitted model), it is convenient to adopt the bayesian interpretation of the smoothing process as applied to the linear taylor expansion considered here.  Following \citet{wahba_1983, silverman_1985, nychka_1988, ruppert_2003}, $\lambda$ can be regarded as the variance on a gaussian prior over the size of $\bm{\theta}$.  If the estimated optimal penalty parameter $\hat\lambda \approx \lambda$, \citet{wahba_1983} found for smoothing splines that bayesian credible bands around some estimated function $\hat{g}(X)$ derived from the penalized spline estimator have good frequentist properties in an ``across-the-function'' sense -- the true function was found within the estimated credible band for $\hat{g}(X)$, approximately at the nominal rate in large samples.  While the individual parameters themselves are biased, this perspective shifts the focus of inference from individual parameters to the smooth functions that they represent.  Given the similarities between splines and neural networks -- both are overspecified nonlinear transformations of a feature vector\footnote{While splines transform a feature vector into a form that can be estimated by a (generalized) linear model, we're concerned here with the linear approximation to a nonlinear model.} -- we proceed from the supposition that this finding may generalize to our case, and explore whether or not it does by simulation in section \ref{simulations_section}.

The standard errors of the parametric terms may thus be calculated in the usual way from the relevant diagonals of $\bm{\hat{V}_\theta}$, while the variance of predictions requires the computation of $\bm{\tilde\xi(Z^*)}$ for new data $\bm{Z^*}$.  Pointwise variances are the diagonal of $\bm{\tilde\xi \hat{V}_\theta \tilde\xi^T}$.  

\section{Simulations}\label{simulations_section}

This section explores finite-sample properties of the proposed estimator through application to a synthetic dataset, and comparison against standard fixed-effects regression techniques.  The synthetic dataset is designed to be intrinsically nonlinear and interactive -- a setting in which neural nets should greatly outperform linear models.

\subsection{Data-generating process}

We generate $n_t$ observations of data $\bm{Z}$ for each of $n_i$ cross-sectional units.  The data $\bm{Z}_{it}$ is distributed multivariate-normally in 5 dimensions, with each cross-sectional unit drawing from its own random mean vector and covariance matrix.  The deterministic outcome $y^*_{it}$ is set equal to $\alpha_i + t + log(\phi(\bm{Z}_{it}))$, where $\phi$ is the standard normal density function in 5 dimensions, $\alpha$ is a fixed effect equal to $i$, and $t$ is time.  As such, the effect of $\bm{Z}$ is nonlinear, the marginal effect of $z\in\bm{Z}$ is dependent on $\bm{Z}_{-z}$, and is correlated with the group-level intercept.  We add noise by setting $y_{it} = y^*_{it}+\mathcal{N}\left(0, 400\right)$.

The data are split into training, testing, and validation sets, by time period.  Observations in the most ``recent'' decile comprise the validation set, which is held aside to gauge out-of-sample performance on the selected model.  The training and testing sets are divided among the remainder from even and odd-numbered time periods.  

\subsection{Fitting}
These data were fit with panel neural nets of the form $y_{it} = \alpha_i + P t + f(\bm{Z}_{it}) + \epsilon$, as well as with standard fixed-effects models of the form $y_{it} = \alpha_i + P t + \bm{Z}_{it}\bm{\beta} + \epsilon$.  In the neural net, the coefficient on the time trend is left unpenalized in order to observe any bias, and to observe properties of their estimated confidence intervals.  The $f(\bm{Z}_{it})$ are represented with a 4-layer network, with 12, 11, 10, and 9 nodes from the bottom to the top.  The activation function $a()$ is the ``leaky'' rectified linear unit \cite{maas_2013};
\[
f(x) = \left\{\begin{array}{l l}
x/100	& \text{if } x<0\\
x	& \text{if } x>0\\
\end{array}\right.
\]  

The penalty parameter $\lambda$ is chosen iteratively.  The simulation commences with a large penalty ($\lambda = 8$), and the parameters are estimated subject to that penalty.  At convergence, the model is saved, and $\lambda_{new} \leftarrow \lambda_{old}/2$.  The model is then re-fit, starting from the parameter values of the previous model, with some random jitter added to help avoid becoming trapped in local minima.

Fitting proceeds with progressively smaller values of $\lambda$ until prediction error on the test set fails to improve for three iterations.  The selected model is the one that minimizes $\E(y_{test} - \hat{y}_{test})^2$.  Note that $k$-fold cross-validation is generally preferable to the training set/test set approach taken here, but would require each model to be refit $k$ times, and thus would be computationally-prohibitive in our monte-carlo setting.

This process is repeated 1000 times for different draws of the DGP.  The selected model is assessed for mean squared error in the validation set, bias in the parametric coefficient, coverage of estimated intervals around the parametric coefficient, and average coverage of estimated intervals around individual predictions in the validation set.  

\subsection{Results}

Results are presented in table \ref{simulations_table}.  The panel neural nets provide substantially better accuracy than do fixed-effects regressions, and provide parametric estimates that are approximately unbiased.  Confidence and prediction intervals -- on parametric estimates and predictions respectively -- cover at approximately nominal rates.  

\begin{table}[htbp]\small
\caption{Results of monte carlo simulations described in section \ref{simulations_section}.  Standard errors in parenthesis.}
\label{simulations_table}
\begin{center}
\begin{tabular}{cccccrrrrr}
\hline
\multicolumn{1}{l}{$T$} & \multicolumn{1}{l}{N. Train} & \multicolumn{1}{l}{N. Test} & \multicolumn{1}{l}{N. Val}  & \multicolumn{1}{l}{MSE} & \multicolumn{1}{l}{MSE (FE)} & \multicolumn{1}{l}{Bias $\hat\beta$} & \multicolumn{1}{l}{$\hat\beta$ coverage} & \multicolumn{1}{l}{$\hat y$ coverage} \\ \hline
\multirow{2}{*}{20} & \multirow{2}{*}{900} & \multirow{2}{*}{900} & \multirow{2}{*}{200}  & 668 & 1212 & -0.0018 & 0.967 & 0.9361 \\ 
&&&&  (88.16) & (175.5) & (0.1549) \\ \hline%& (0.1787) & (0.003162) \\ \hline
\multirow{2}{*}{40} & \multirow{2}{*}{900} & \multirow{2}{*}{900} & \multirow{2}{*}{200}  & 627.1 & 1165 & 0.001776 & 0.987 & 0.9498 \\ 
&&&&  (84.6) & (179.3) & (0.07569) & \\ \hline%(0.1134) & (0.002889) \\ \hline
\multirow{2}{*}{20} & \multirow{2}{*}{1800} & \multirow{2}{*}{1800} & \multirow{2}{*}{400}  & 667.5 & 1227 & 0.003283 & 0.958 & 0.9288 \\ 
&&&  & (68.96) & (128.1) & (0.1113) &\\ \hline% (0.2007) & (0.03172) \\

\multirow{2}{*}{40} & \multirow{2}{*}{1800} & \multirow{2}{*}{1800} & \multirow{2}{*}{400}  & 626 & 1182 & 0.0002555 & 0.958 & 0.9429 \\ 
&&& & (64.12) & (127.4) & (0.05525) & \\ \hline%(0.2007) & (0.02991) \\ 

\multirow{2}{*}{20} & \multirow{2}{*}{2700} & \multirow{2}{*}{2700} & \multirow{2}{*}{600}  & 664.1 & 1225 & -0.0005533 & 0.95 & 0.9268 \\ 
&& & & (58.8) & (99.36) & (0.08988) & \\ \hline%(0.2181) & (0.03172) \\
\multirow{2}{*}{40} & \multirow{2}{*}{2700} & \multirow{2}{*}{2700} & \multirow{2}{*}{600}  & 624.6 & 1191 & 0.00001122 & 0.953 & 0.9416 \\ 
&&& & (57.77) & (111.4) & (0.2117) &\\ \hline% (0.216) & (0.02991) \\ 
\hline
\end{tabular}
\end{center}
\end{table}

\FloatBarrier
\section{Application:  predicting county-level agricultural yields from weather}

This final section applies the panel neural net to the prediction of agricultural yields from weather data.  This problem is relevant for short-term economic forecasts as well as for longer-range climate change impact assessment.  

It is also a natural use-case for a semi-parametric panel data model.  First, weather data over an entire year is high-dimensional.  While there is only one outcome per year in our setting, the explanatory variables come in the form of several weather variables over hundreds of days of the growing season.  Models that penalize high-dimensional covariate sets can help avoid overfitting.  

Second, additively-separable models like ridge regression or LASSO may not be appropriate.  The ``law of the minimum'' -- a generally-accepted concept in ecology and agronomy -- states that plant growth is limited by whatever single growth factor is most constrained.  We therefore require an inherently nonlinear and interactive representation of the process that maps weather to yields.

Third, machine-learning algorithms that do not have a linear representation -- like decision trees and random forests -- have no clear way to incorporate unobserved cross-sectional heterogeneity\footnote{One exception is the RE-EM trees of \citet{sela_2012}, which can incorporate random effects into decision tree estimation.}, which can be important -- especially in linear fixed-effect models that do not admit time-invariant regressors.

Finally, agricultural yields in developed countries have seen substantial increases in recent years, driven primarily by technological change.  This is naturally represented by a parametric trend, which is incompatible with models like random forests that cannot extrapolate beyond the support of the training data.

\subsection{Data}

County-level corn yield data for Iowa, Illinois, and Missouri is taken from the National Agricultural Statistical Service of the US Department of Agriculture.  We use only county-years with at least 5000 acres planted in corn, and only counties with at least 20 such years.  We exclude all counties that have more than 20\% of their agricultural land under irrigation.  We focus on grain corn, rather than silage.  

Weather data comes from PRISM \citep{PRISM}.  From its native gridded format, daily data is aggregated into areas corresponding to the agricultural area of each county.  We posses measurements of minimum and maximum temperature, precipitation, and dewpoint.  These data are subset into a growing season that runs from April through October, leaving us with a total of 856 weather variables.

In addition, we augment the weather data with (time-invariant) soils data from SSURGO \citep{SSURGO}.  The subset we use contains measurements of 38 chemical and physical soil properties, averaged over the county.  While such data in linear fixed-effects models will generally be collinear with the individual effects and thereby dropped, the nonlinear nature of the panel neural net allows these variables to moderate the effects of the time-varying weather data.  We also include latitude and longitude as regressors, which allows estimated relationships to vary smoothly in space.

\subsection{Models}

We model these data from the period 1981 to 2014, saving 2015 as a test set for evaluation of our fitted models.  These are:

\subsubsection{FE-OLS}

A standard fixed effects model, we fit
\[
y_{it} = \alpha_i + \left[\text{State}_i\times[time_t, time_t^2]\right]\bm\beta + \bm{Z}_{it}\bm{\Gamma} + \epsilon_{it}
\]

\noindent Note that time-invariant soil and location data is collinear with the fixed effects, and dropped from $\bm{Z}$.  

\subsubsection{LASSO}

As above, but solving for $\bm\theta = [\bm{\beta, \Gamma}]$ to minimize $\Sigma(y-\hat{y}) + \lambda\Sigma|\theta|$.  

\subsubsection{Random Forest}

The model is 
\[
y_{it} = g\left(\left[\text{State}_i\times[time_t]\right], \bm{Z}_{it}\right)+ \epsilon_{it}
\]
\noindent and as such does not incorporate any information on the longitudinal nature of the data, except the time trends. We grow 500 trees, with each split in each tree selecting from 1/3rd of the available variables.

\subsubsection{Panel neural network}
The models is
\[
y_{it} = \alpha_i + \left[\text{State}_i\times[time_t, time_t^2]\right]\bm\beta + g\left(\bm{Z}_{it}\right) + \epsilon_{it}
\]
The quadratic state-by-year time trends are subject to the weight decay penalty, along with $\bm{\Gamma}$.  We use an architecture of 40 nodes on the bottom layer, 20 in a middle layer, and 10 in a top layer.  The activation function is the rectified linear unit, and training is by batch gradient descent using the adaptive step size algorithm (``RMSprop'') of \citet{tieleman_2012}.

\subsubsection{Results}
\FloatBarrier
\begin{figure}[htbp]
\caption{Prediction results}
\label{maps_results}
\begin{center}
\includegraphics[scale = .8]{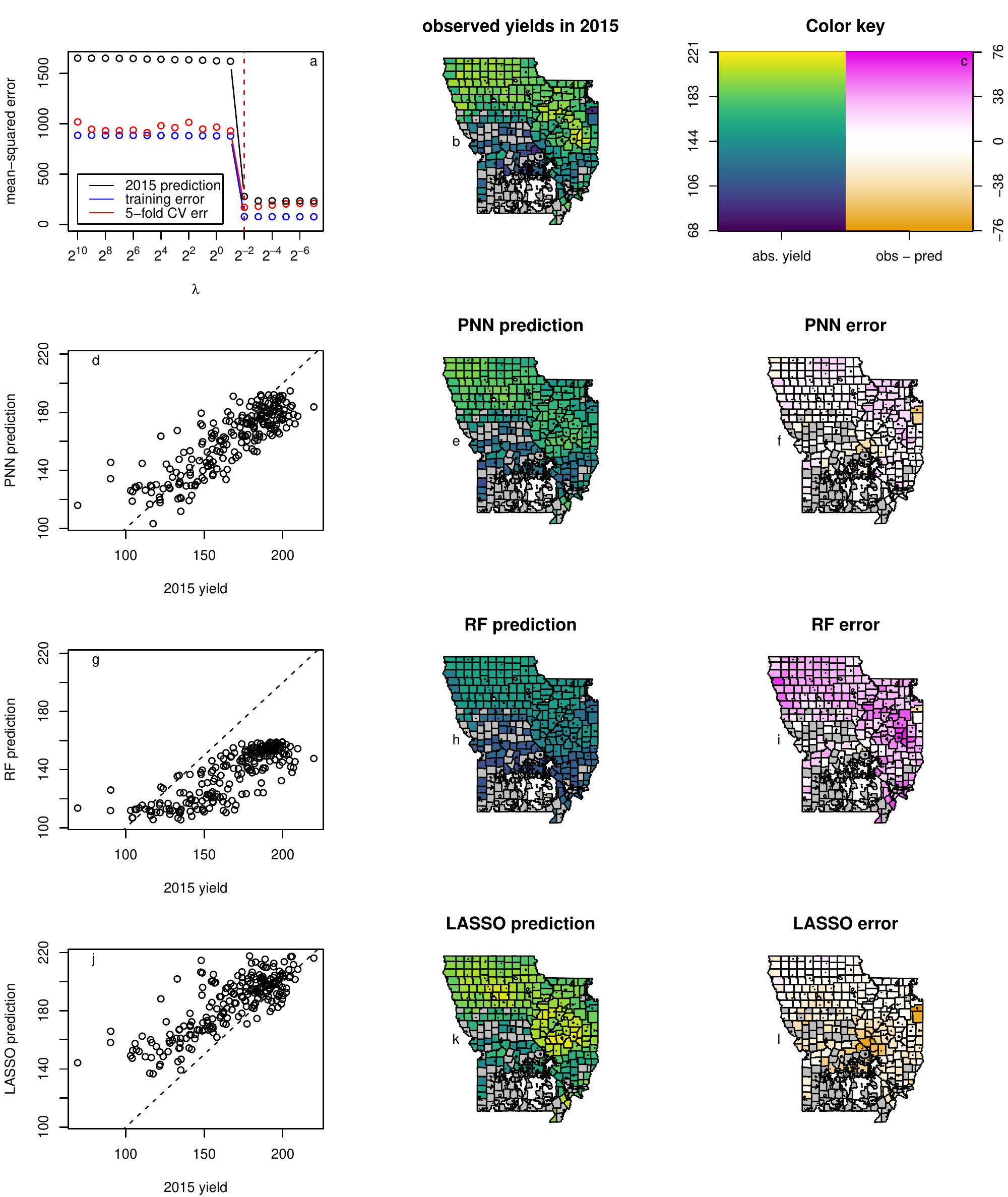}
\end{center}
\end{figure}

Results are presented in figure \ref{maps_results}.  Figure \ref{maps_results}a shows in-sample, cross-validation, and 2015 prediction errors for the panel neural net -- 5-fold cross-validation predicted the 2nd best value of $\lambda$ out of those tested.  For this model, mean-squared error was 273, and predictions were approximately unbiased (figure \ref{maps_results}d), though some under-prediction is apparent in north-central Illinois (\ref{maps_results}f).  

The random forest's MSE was 1197, and its poor performance appears to owe to substantial bias (\ref{maps_results}g).  Indeed, its rampant under-prediction appears to stem from its inability to model a linear time trend such that it can effectively extrapolate one year beyond its training sample -- operating like a nearest-neighbor algorithm, random forest will model 2015 as being similar to 2014 and adjacent years, and have no way of discerning a long-term trend in a noisy signal.

The LASSO performs better here than the random forest, with a MSE of 609.  Its predictions are biased slightly upwards.  This likely stems from over-reliance on the time trend, as this model was unable to discern nonlinear functions the daily training data.

The fixed-effects model is not shown, as the optimal value of $\lambda$ for the lasso was $0.015$ -- quite close to zero and thus to the OLS solution.

\FloatBarrier
\section{Conclusions and Future Work}

This paper has proposed an adaptation of neural networks to panel data, and demonstrated (1) unbiasedness of parametric estimates, (2) good properties of estimated confidence intervals, and (3) efficiency both in a simulated dataset and in an application to yield prediction from weather data.

Further work will provide refinements to the \texttt{panelNNET} software used, and demonstrate the applicability of this class of models to the problem of heterogeneous treatment effect estimation from randomized and quasi-experiments.

\bibliographystyle{unsrtnat}
\bibliography{citations.bib}

\begin{thebibliography}{31}
\providecommand{\natexlab}[1]{#1}
\providecommand{\url}[1]{\texttt{#1}}
\expandafter\ifx\csname urlstyle\endcsname\relax
  \providecommand{\doi}[1]{doi: #1}\else
  \providecommand{\doi}{doi: \begingroup \urlstyle{rm}\Url}\fi

\bibitem[Rosenblatt(1958)]{rosenblatt_1958}
Frank Rosenblatt.
\newblock The perceptron: a probabilistic model for information storage and
  organization in the brain.
\newblock \emph{Psychological review}, 65\penalty0 (6):\penalty0 386, 1958.

\bibitem[LeCun et~al.(2015)LeCun, Bengio, and Hinton]{lecun_2015}
Yann LeCun, Yoshua Bengio, and Geoffrey Hinton.
\newblock Deep learning.
\newblock \emph{Nature}, 521\penalty0 (7553):\penalty0 436--444, 2015.

\bibitem[Breiman(2001)]{breiman_2001}
Leo Breiman.
\newblock Random forests.
\newblock \emph{Machine learning}, 45\penalty0 (1):\penalty0 5--32, 2001.

\bibitem[Wager et~al.(2014)Wager, Hastie, and Efron]{wager_2014}
Stefan Wager, Trevor Hastie, and Bradley Efron.
\newblock Confidence intervals for random forests: the jackknife and the
  infinitesimal jackknife.
\newblock \emph{Journal of Machine Learning Research}, 15\penalty0
  (1):\penalty0 1625--1651, 2014.

\bibitem[Wager and Athey(2015)]{wager_2015}
Stefan Wager and Susan Athey.
\newblock Estimation and inference of heterogeneous treatment effects using
  random forests.
\newblock \emph{arXiv preprint arXiv:1510.04342}, 2015.

\bibitem[Zou and Hastie(2005)]{zou_2005}
Hui Zou and Trevor Hastie.
\newblock Regularization and variable selection via the elastic net.
\newblock \emph{Journal of the Royal Statistical Society: Series B (Statistical
  Methodology)}, 67\penalty0 (2):\penalty0 301--320, 2005.

\bibitem[Hastie and Tibshirani(1990)]{hastie_1990}
Trevor~J Hastie and Robert~J Tibshirani.
\newblock \emph{Generalized additive models}, volume~43.
\newblock CRC Press, 1990.

\bibitem[Henderson et~al.(2008)Henderson, Carroll, and Li]{henderson_2008}
Daniel~J Henderson, Raymond~J Carroll, and Qi~Li.
\newblock Nonparametric estimation and testing of fixed effects panel data
  models.
\newblock \emph{Journal of Econometrics}, 144\penalty0 (1):\penalty0 257--275,
  2008.

\bibitem[Hoderlein and White(2012)]{hoderlein_2012}
Stefan Hoderlein and Halbert White.
\newblock Nonparametric identification in nonseparable panel data models with
  generalized fixed effects.
\newblock \emph{Journal of Econometrics}, 168\penalty0 (2):\penalty0 300--314,
  2012.

\bibitem[Li and Liang(2015)]{li_2015}
Cong Li and Zhongwen Liang.
\newblock Asymptotics for nonparametric and semiparametric fixed effects panel
  models.
\newblock \emph{Journal of Econometrics}, 185\penalty0 (2):\penalty0 420--434,
  2015.

\bibitem[Chernozhukov et~al.(2016)Chernozhukov, Chetverikov, Demirer, Duflo,
  Hansen, et~al.]{chernozhukov_2016}
Victor Chernozhukov, Denis Chetverikov, Mert Demirer, Esther Duflo, Christian
  Hansen, et~al.
\newblock Double machine learning for treatment and causal parameters.
\newblock \emph{arXiv preprint arXiv:1608.00060}, 2016.

\bibitem[Athey and Imbens(2016)]{athey_2016_trees}
Susan Athey and Guido Imbens.
\newblock Recursive partitioning for heterogeneous causal effects.
\newblock \emph{Proceedings of the National Academy of Sciences}, 113\penalty0
  (27):\penalty0 7353--7360, 2016.

\bibitem[Nair and Hinton(2010)]{nair_2010}
Vinod Nair and Geoffrey~E Hinton.
\newblock Rectified linear units improve restricted boltzmann machines.
\newblock In \emph{Proceedings of the 27th International Conference on Machine
  Learning (ICML-10)}, pages 807--814, 2010.

\bibitem[Maas et~al.(2013)Maas, Hannun, and Ng]{maas_2013}
Andrew~L Maas, Awni~Y Hannun, and Andrew~Y Ng.
\newblock Rectifier nonlinearities improve neural network acoustic models.
\newblock In \emph{Proc. ICML}, volume~30, 2013.

\bibitem[He et~al.(2015)He, Zhang, Ren, and Sun]{he_2015}
Kaiming He, Xiangyu Zhang, Shaoqing Ren, and Jian Sun.
\newblock Delving deep into rectifiers: Surpassing human-level performance on
  imagenet classification.
\newblock In \emph{Proceedings of the IEEE international conference on computer
  vision}, pages 1026--1034, 2015.

\bibitem[Hornik et~al.(1989)Hornik, Stinchcombe, and White]{hornik_1989}
Kurt Hornik, Maxwell Stinchcombe, and Halbert White.
\newblock Multilayer feedforward networks are universal approximators.
\newblock \emph{Neural networks}, 2\penalty0 (5):\penalty0 359--366, 1989.

\bibitem[Friedman et~al.(2001)Friedman, Hastie, and Tibshirani]{friedman_2001}
Jerome Friedman, Trevor Hastie, and Robert Tibshirani.
\newblock \emph{The elements of statistical learning}, volume~1.
\newblock Springer series in statistics Springer, Berlin, 2001.

\bibitem[Ruder(2016)]{ruder_2016}
Sebastian Ruder.
\newblock Gradient descent variants.
\newblock 2016.

\bibitem[Rivals and Personnaz(2000)]{rivals_2000}
Isabelle Rivals and L{\'e}on Personnaz.
\newblock Construction of confidence intervals for neural networks based on
  least squares estimation.
\newblock \emph{Neural Networks}, 13\penalty0 (4):\penalty0 463--484, 2000.

\bibitem[Cameron et~al.(2012)Cameron, Gelbach, and Miller]{cameron_2012}
A~Colin Cameron, Jonah~B Gelbach, and Douglas~L Miller.
\newblock Robust inference with multiway clustering.
\newblock \emph{Journal of Business \& Economic Statistics}, 2012.

\bibitem[Wahba(1983)]{wahba_1983}
Grace Wahba.
\newblock Bayesian" confidence intervals" for the cross-validated smoothing
  spline.
\newblock \emph{Journal of the Royal Statistical Society. Series B
  (Methodological)}, pages 133--150, 1983.

\bibitem[Silverman(1985)]{silverman_1985}
Bernhard~W Silverman.
\newblock Some aspects of the spline smoothing approach to non-parametric
  regression curve fitting.
\newblock \emph{Journal of the Royal Statistical Society. Series B
  (Methodological)}, pages 1--52, 1985.

\bibitem[Nychka(1988)]{nychka_1988}
Douglas Nychka.
\newblock Bayesian confidence intervals for smoothing splines.
\newblock \emph{Journal of the American Statistical Association}, 83\penalty0
  (404):\penalty0 1134--1143, 1988.

\bibitem[Ruppert et~al.(2003)Ruppert, Wand, and Carroll]{ruppert_2003}
David Ruppert, Matt~P Wand, and Raymond~J Carroll.
\newblock \emph{Semiparametric regression}.
\newblock Number~12. Cambridge university press, 2003.

\bibitem[Sela and Simonoff(2012)]{sela_2012}
Rebecca~J Sela and Jeffrey~S Simonoff.
\newblock Re-em trees: a data mining approach for longitudinal and clustered
  data.
\newblock \emph{Machine learning}, 86\penalty0 (2):\penalty0 169--207, 2012.

\bibitem[Hart and Bell(2015)]{PRISM}
Edmund~M. Hart and Kendon Bell.
\newblock \emph{prism: Download data from the Oregon prism project}, 2015.
\newblock URL \url{http://github.com/ropensci/prism}.
\newblock R package version 0.0.6.

\bibitem[Soil Survey~Staff(2016)]{SSURGO}
United States Department of~Agriculture. Soil Survey~Staff, Natural Resources
  Conservation~Service.
\newblock \emph{Soil Survey Geographic (SSURGO) Database.}, 2016.
\newblock URL \url{https://sdmdataaccess.sc.egov.usda.gov}.

\bibitem[Tieleman and Hinton(2012)]{tieleman_2012}
Tijmen Tieleman and Geoffrey Hinton.
\newblock Lecture 6.5-rmsprop: Divide the gradient by a running average of its
  recent magnitude.
\newblock \emph{COURSERA: Neural networks for machine learning}, 4\penalty0
  (2), 2012.

\bibitem[Krizhevsky et~al.(2012)Krizhevsky, Sutskever, and
  Hinton]{krizhevsky_2012}
Alex Krizhevsky, Ilya Sutskever, and Geoffrey~E Hinton.
\newblock Imagenet classification with deep convolutional neural networks.
\newblock In \emph{Advances in neural information processing systems}, pages
  1097--1105, 2012.

\bibitem[Schlenker and Roberts(2009)]{schlenker_2009}
Wolfram Schlenker and Michael~J Roberts.
\newblock Nonlinear temperature effects indicate severe damages to us crop
  yields under climate change.
\newblock \emph{Proceedings of the National Academy of sciences}, 106\penalty0
  (37):\penalty0 15594--15598, 2009.

\bibitem[Bengio et~al.(2007)Bengio, Lamblin, Popovici, Larochelle,
  et~al.]{bengio_2007}
Yoshua Bengio, Pascal Lamblin, Dan Popovici, Hugo Larochelle, et~al.
\newblock Greedy layer-wise training of deep networks.
\newblock \emph{Advances in neural information processing systems},
  19:\penalty0 153, 2007.

\end{thebibliography}

\FloatBarrier

\end{document}